\documentclass[usenatbib,usegraphicx]{mn2e}

\def\apj{ApJ}
\def\apjs{ApJS}
\def\aap{A\&A}

\def\aaps{A\&AS}
\def\mnras{MNRAS}
\def\aj{AJ}
\def\nat{Nature}
\def\araa{ARA\&A}
\def\pasp{PASP}

\def\apss{Ap\&SS}

\def\cm{\textrm{cm}}
\def\meter{\textrm{m}}
\def\km{\textrm{km}}

\def\mag{\textrm{mag}}
\def\arcsec{\textrm{arcsec}}
\def\arcmin{\textrm{arcmin}}
\def\mm{\textrm{mm}}
\def\msec{\textrm{ms}}
\def\Ang{\textrm{\AA}}

\title[Cherenkov telescopes as ELTs]{Cherenkov Telescopes as Optical Telescopes for Bright Sources: Today's Specialised Thirty Metre Telescopes?}
\author[Lacki]{Brian C. Lacki$^{1,2}$\thanks{E-mail:lacki@astronomy.ohio-state.edu}\\
$^1$Department of Astronomy, The Ohio State University, 140 West 18th Avenue, Columbus, OH 43210, USA\\
$^2$Center for Cosmology \& Astro-Particle Physics, The Ohio State University, Columbus, Ohio 43210, USA}

\begin{document}

\maketitle

\begin{abstract}
Imaging Atmospheric Cherenkov Telescopes (IACTs) use large-aperture ($\sim 10 - 30\ \meter$) optical telescopes with arcminute angular resolution to detect TeV gamma-rays in the atmosphere.  I show that IACTs are well-suited for optical observations of bright sources ($V \la 8 - 10$), because these sources are brighter than the sky background.  Their advantages are especially great on rapid time-scales.  Thus, IACTs are ideal for studying many phenomena optically, including transiting exoplanets and the brightest gamma-ray bursts.  In principle, an IACT could achieve millimagnitude photometry of these objects with second-long exposures.  I also consider the potential for optical spectroscopy with IACTs, finding that their poor angular resolution limits their usefulness for high spectral resolutions, unless complex instruments are developed.  The high photon collection rate of IACTs is potentially useful for precise polarimetry.  Finally, I briefly discuss the broader possibilities of extremely large, low resolution telescopes, including a 10 arcsec~resolution telescope and spaceborne telescopes.
\end{abstract}

\begin{keywords}
telescopes -- techniques: photometric -- techniques -- spectroscopic -- techniques: polarimetric
\end{keywords}

\section{Introduction}
\label{sec:Introduction}

Extremely Large Telescopes (ELTs) with apertures of $30\ \meter$ promise to open a new era in astronomy, but will be extremely costly and complex.  It is less widely appreciated that we \emph{already} have ELT-size telescopes of low optical resolution: the Imaging Atmospheric Cherenkov Telescopes (IACTs).  IACTs are designed primarily to image TeV gamma-ray sources by detecting the Cherenkov radiation from the particle showers created when a TeV gamma-ray hits the atmosphere \citep[for reviews of the uses of IACTs as TeV detectors, see for example][]{Aharonian97,Ong98,Aharonian08}.  Detecting the faint Cherenkov pulse requires high photon statistics, and reconstructing the structure and direction of the shower is aided by stereoscopic imaging.  IACTs therefore typically consist of an array of multiple mirrors, each with large collecting areas ($\sim 100\ \meter^2$).  Many IACTs have been built, including HESS \citep{Bernlohr03,Cornils03}, VERITAS \citep{Weekes02,Humensky07}, MAGIC \citep{Lorenz04}, CANGAROO \citep{Kawachi01}, HEGRA \citep{Akhperjanian98}, and Whipple \citep{Lewis90,Kildea07}.  Future IACT arrays like the Cherenkov Telescope Array (CTA) may have four $25\ \meter$ mirrors and tens of $10\ \meter$ mirrors, for an optical collection area equivalent to a $\ga 50\ \meter$ telescope \citep{Doro09}.  The cost of the IACT mirrors is very low, only $\sim \pounds 2000 / \meter^2\ (2010)$, and may go down further in the future \citep{CTA10}.  However, the lower quality IACTs have poor angular resolution by optical standards ($\sim 1 - 2'$; \citealt{Bernlohr03,Cornils03}).  The huge size of the mirrors still makes them ideal `light buckets', and the primary science cannot occur on moonlit nights.  IACTs have therefore been used for some optical studies, often when the sky glow is brighter: detecting the optical pulsations of the Crab pulsar \citep{OnaWilhelmi04,Hinton06,Lucarelli08}, searches for microsecond optical transients \citep{Deil09}, and even optical SETI \citep{Eichler01,Holder05}.  IACTs have also been proposed as tools for stellar intensity interferometry \citep{LeBohec06}\footnote{Conversely, the Narrabri Observatory intensity interferometer, which like an IACT array used several low angular resolution reflectors \citep*{HanburyBrown67}, carried out one of the earliest Southern Hemisphere searches for TeV gamma-rays (\citealt*{HanburyBrown69}; \citealt{Grindlay75}).} and detecting nanosecond optical transients \citep{Borra10}.  However, the potential of IACTs to study bright, steady sources other than pulsars has not been studied in great detail.  

The primary challenge to using IACTs as optical telescopes is that their low angular resolution blends sky background with a target source's light.  When the source is brighter than the sky background, the IACT will be at its most effective.  On a moonlit night, the {\emph V}-band sky brightness is $\mu_V \approx 20\ \mag\ \arcsec^{-2}$ (depending on the phase and Moon--source angle; \citealt{Krisciunas91}; see also Fig. 4 of \citealt{Deil09}), and the sky brightness within the PSF is  
\begin{equation}
\label{eqn:VSky}
V_{\rm sky} \approx 8.36 - 5.0\ \log_{10} \left(\frac{\theta_{\rm PSF}}{2\ \arcmin}\right)
\end{equation}
where $\theta_{\rm PSF} \approx 1 -- 2'$ are typical radii of PSFs for IACTs.  Similar values are found for B and I bands.  On moonless nights, the {\emph V}-band sky brightness at IACT sites is typically $22\ \mag\ \arcsec^{-2}$ to give $V_{\rm sky} \approx 10.4$, at $|b| \ge 20^{\circ}$ and including all stars with $V \le 6$ \citep{Preu02}.  Confusion will also limit the usefulness of IACTs for stars with $V \ga 14 - 16$ ($|b| \ge 20^{\circ}$), as the images of the stars begin to overlap \citep{Bahcall80}.  

\begin{figure}
\centerline{\includegraphics[width=8cm]{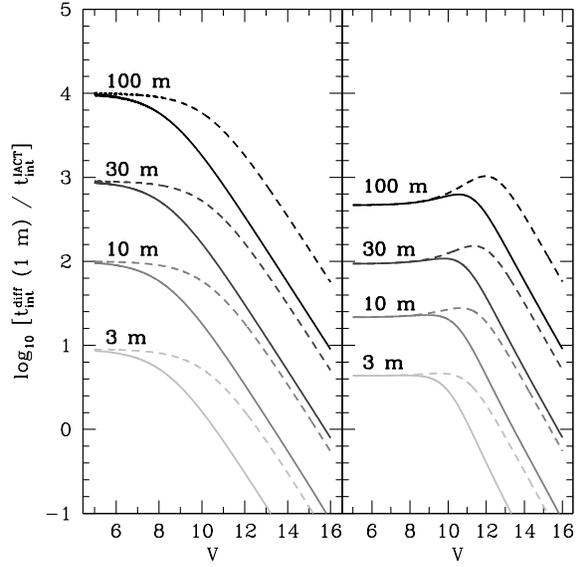}}
\caption{Relative exposure time with a diffraction limited 1 metre telescope needed to achieve the same $S/N$ as a $2\ \arcmin$ resolution IACT, as a function of {\emph V}-band magnitude.  On left, only Poisson noise is considered (as appropriate for spectroscopy and polarimetry); on right, scintillation noise (with airmass $X = 1$) is taken into account (as appropriate for photometry).  Both moonlit (solid) and moonless (dashed) nights are plotted. \label{fig:tIntRatio}}
\end{figure}

The point of this paper is to explore whether and how IACTs can exploit their huge photon collection rates for bright source optical astronomy.  During a {\emph V}-band ($\lambda = 5448\ \Ang$; $\Delta\lambda = 840\ \Ang$; \citealt{Bessell05}) observation, IACTs collect
\begin{equation}
\label{eqn:NphStar}
N_{\star} \approx 2.7 \times 10^8 \times 10^{-0.4 (V - 8)} A_{100} t_{\rm int} \eta_{0.5}
\end{equation}
photons, where $A_{100} 100\ \meter^2$ is the collecting area, $0.5 \eta_{0.5}$ is the photon detection efficiency, and $t_{\rm int}$ is the integration time in seconds.  IACTs are most suited for high speed measurements, where brighter objects will be the detectable ones in any telescope of a given aperture \citep{Deil09}.  Fig.~\ref{fig:tIntRatio} shows that the integration times relative to a 1 metre diffraction-limited telescope.  With bright sources, IACTs can achieve the same signal-to-noise with integration times that are orders of magnitude shorter.   IACTs also are better when the sky background is low, particularly in bluer sky bands on moonless nights.  We now consider in turn the potentials and challenges of photometry (\S~\ref{sec:Photometry}), spectroscopy (\S~\ref{sec:Spectroscopy}), and polarimetry (\S~\ref{sec:Polarimetry}) with IACTs.  

\section{Photometry with IACTs}
\label{sec:Photometry}
Photometry is limited by several kinds of noise: detector noise, which I mostly ignore here; Poisson fluctuations from the source and the sky background; scintillation noise; and systematic red noise (\citealt*{Pont06}; see \S~\ref{sec:Systematics}).  The Poisson fluctuations from the sky and source is given by $S/N = N_{\star} / \sqrt{N_{\star} + N_{\rm sky}}$ where $N_{\star}$ is the number of photons from the target source and $N_{\rm sky}$ is the number of photons from the sky blended with the source image.  When $S/N \gg 1$, the noise in the magnitude is $\sigma^{\rm Poisson}_V \approx 1.0857 (S/N)^{-1}$.  In the bright source regime ($V \ll V_{\rm sky}$), we therefore have $S/N \approx \sqrt{N_{\star}}$, or 
\begin{equation}
\label{eqn:sigmaVPoissonBright}
\sigma^{\rm Poisson}_V \approx 6.7 \times 10^{-5} \times 10^{0.2 (V - 8)} / \sqrt{A_{100} t_{\rm int} \eta_{0.5}}.
\end{equation}
When the source is faint ($V \gg V_{\rm sky}$), $S/N \approx N_{\star} / \sqrt{N_{\rm sky}}$, or
\begin{equation}
\label{eqn:sigmaVPoissonDim}
\sigma^{\rm Poisson}_V \approx 5.6 \times 10^{-5} \times 10^{0.2 (2V - 16 - V_{\rm sky} + 8.36)} / \sqrt{A_{100} t_{\rm int} \eta_{0.5}}.
\end{equation}
Finally there is scintillation noise caused by turbulence in the atmosphere, which is given in \citet{Young67} and \citet{Southworth09}:
\begin{equation}
\label{eqn:sigmaVScint}
\sigma^{\rm scint}_V = 5.6 \times 10^{-4} A_{100}^{-1/3} t_{\rm int}^{-1/2} X^{7/4} e^{-h / 8\ {\rm km}}
\end{equation}
where $X$ is the airmass and $h$ is the altitude of the telescope, and is typically $\sim 2\ \km$ \citep{Preu02}.  For exposure times less than $10\ \msec$, the spectral index of atmospheric turbulence changes and the scintillation noise becomes even greater, down to $0.3\ \msec$ when the inner scale of turbulence is reached \citep{Dravins97}.  On the other hand, IACTs may be large enough to be affected by the outer scale of turbulence \citep{Dravins97b}, which ranges somewhere between a few metres \citep[e.g.,][]{Nightingale91} and several kilometres \citep*[e.g.,][]{Colavita87}, in which case the scintillation noise may decrease \citep{Dravins98}.  The prospects for probing the outer scale are especially good if multiple telescopes in an array like HESS or VERITAS are used, since they are typically tens of metres apart.

In practice, the sky background noise may be greater if the image plane is not focused to infinity.  IACTs are focused to the atmosphere, which spreads out the PSF of sources at infinity and dilutes their light with the sky background \citep[e.g.,][]{Lucarelli08}.  For a reflector of focal length $f$ with light from a distance $s$ in focus on the detector plane, the detector plane would have to be shifted by a length $\Delta s' \approx f^2/s \approx 1\ \cm (f / 10\ \meter)^2 (s / 10\ \km)^{-1}$ for sources at infinity to be in focus \citep[e.g.,][]{Schroeder00}.  Otherwise, the PSF radius will be broadened by an apparent angle $\theta_{\rm focus} \approx D / (2 s) \approx 1\farcm7 (D / 10\ \meter) (s / 10\ \km)^{-1}$ for a telescope of aperture $D$.  \citet{Deil09} used a secondary mirror to put distant sources in focus on a custom detector, so focusing is not an insurmountable issue.  

\begin{figure}
\centerline{\includegraphics[width=8cm]{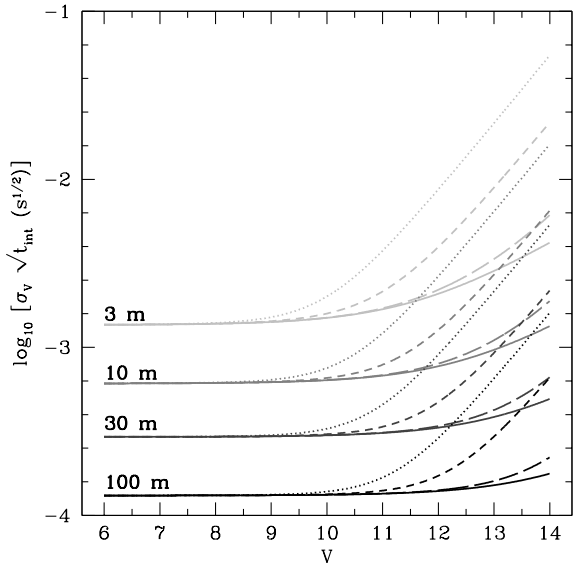}}
\caption{Photometric white noise $\sigma_V$ achieved by ground-based telescopes with different apertures and PSF sizes, as a function of $V$ and $t_{\rm int}$.  Considered apertures are 3 m, 10 m, 30 m, and 100 m; sky backgrounds are calculated assuming moonlit nights with $2\ \arcmin$ (dotted), $10\ \arcsec$ (long-dashed), and diffraction-limited (solid) PSFs.  IACTs observing on moonless nights with $2\ \arcmin$ PSFs are shown as short-dashed lines.  IACTs are well-suited for photometry of bright objects.  Scintillation noise (calculated with $X = 1$) dominates for bright objects.  Note that systematic effects likely exceeds the plotted noise for $\sigma_V \la 0.001$.  For very short $t_{\rm int} \la 0.01~\sec$, scintillation noise may be different than plotted here.\label{fig:sigmaV}}
\end{figure}

The total white noise in the magnitude is given by $\sigma_V = \sqrt{(\sigma^{\rm Poisson}_V)^2 + (\sigma^{\rm scint}_V)^2}$.  I show $\sigma_V$ for IACTs as compared to diffraction-limited and 10 arcsec resolution telescopes in Fig.~\ref{fig:sigmaV}, as a function of V magnitude.  We see that scintillation typically dominates $\sigma_V$ except for faint sources ($V \ga 10 - 14$).  A 10 metre IACT in principle achieves millimagnitude photometry within a few seconds, for $V \la 12$.  For sources with $V \ga V_{\rm sky}$, IACTs are less efficient than high resolution telescopes, because the S/N for IACTs goes as $\sqrt{N_{\star}} \times \sqrt{N_{\star} / N_{\rm sky}}$ while for high resolution telescopes the signal-to-noise goes as $\sqrt{N_{\star}}$.  

\subsection{Detectors}
IACTs use photomultiplier tubes (PMTs) as their optical detectors, and it would be easiest to simply use these for optical observations.  Such observations have been demonstrated with IACT observations of pulsars \citep{Lucarelli08}.  The main advantage of PMTs, aside from already being installed, is that they can monitor photon flux on extremely short time scales, taking full advantage of the IACT's ability to collect photons quickly.  Automated Photoelectric Telescopes have for many years demonstrated the potential of millimagnitude photometry with PMTs \citep[e.g.,][]{Henry95,Henry99}, including one of the first detections of a transiting exoplanet \citep{Henry00}.  Disadvantages of PMTs include lower quantum efficiency, and larger pixel sizes (typically $\sim 0.1^{\circ} - 0.2^{\circ}$; \citealt{Bernlohr03,Lorenz04}) with the currently installed PMTs.  The low PMT angular resolution increases the sky background even further, reducing $V_{\rm sky}$ to $\sim 6$ on moonlit nights and $\sim 8$ on moonless nights for a pixel size of $0.1^{\circ}$.    

CCDs, by contrast, have higher quantum efficiency and much higher spatial resolution.  However, using CCDs would require replacing the current PMT detectors of IACTs\footnote{Unless the CCDs replaced only a small number of PMTs, this would end an IACT's usefulness for TeV astronomy, because the short integration times of PMTs are necessary for discerning the Cherenkov pulse from particle showers.  Another possibility might be to mount the CCDs directly on the PMT camera's lid.  In fact, images of stars on the PMT camera lid are used to align HESS' mirror facets \citep{Cornils03}.  Then a second lid for the CCDs themselves is probably necessary to protect them in turn from the elements.}, or the construction of new IACT-like telescopes designed to use CCDs.  The image of a bright object will be spread over a very large area in an IACT; given a typical focal length of $\sim 15\ \meter$ (image scale of $4.4\ \mm~\arcmin^{-1}$), the image will have a diameter of $18\ \mm\ (\theta_{\rm PSF} / 2\ \arcmin)$, roughly the size of a typical CCD detector.  The large size essentially eliminates flat-fielding noise, but two issues may concern us when using CCDs: read noise and CCD saturation.  For a typical pixel size of $15\ \mu m$, the image will be spread out over $n_{\rm pix} \approx 10^6 (\theta_{\rm PSF} / 2\ \arcmin)^2$ pixels.  From equation~\ref{eqn:NphStar} we expect $\sim 300 \sec^{-1} t_{\rm int} \eta_{0.5}$ photons per pixel, far below saturation for most CCDs for 1 second exposures (note that for a given f-ratio, the image scale for a larger telescope will become proportionally bigger).  Thus, whereas stopped apertures are used for photometry of bright stars with standard telescopes \citep{LopezMorales06}, IACTs evade the saturation problems precisely \emph{because} they have such low angular resolution.  Finally the read noise is $N_{\rm RN} = n_{\rm RN} \sqrt{n_{\rm pix}}$, where $n_{\rm RN} \approx 5$ is the read-out noise per pixel.  The read-out noise is tiny except for the shortest exposures: $\sigma^{\rm RN}_V \approx 1.0857 (N_{\rm RN} / N_{\star}) \approx 2 \times 10^{-5} \times 10^{-0.4 (V - 8)} (\theta_{\rm PSF} / 2\ \arcmin) [A_{100} t_{\rm int} \eta_{0.5}]^{-1}$.  

A major problem with typical CCDs is that they have readout times of $\sim 10 - 100\ \sec$.  Thus, exposure times are typically limited to at least about a minute, since low duty cycles would negate the light gathering power of the telescope.  There are high speed CCD systems, but these typically work by using only a small part of the chip to image a target source, and shifting the image around on the CCD to get multiple exposures on one chip \citep[e.g.,][]{ODonoghue95,Dhillon07}.  This will not be possible with IACTs, as the image of the source will be comparable to the size of the chip.  

Geiger-mode Avalanche Photodiodes (G-APDs; also called silicon photomultipliers or SiPMs) are a relatively new detector system that are now being applied to Cherenkov telescopes \citep[e.g.,][]{Buzhan03,Renker07}.  G-APDs are $\sim$millimetre size detectors consisting of thousands of micro-pixels; the micro-pixels are connected by a substrate, which gives a signal for the detector.  They are extremely good at timing photons ($< 0.1$ ns timing precision), while being able to tolerate high sky backgrounds of GHz per detector \citep[e.g.,][]{Anderhub09} (compare eqn.~\ref{eqn:NphStar}).  Because the individual detectors are smaller than the PMTs currently used in Cherenkov telescopes, G-APDs would not blend the source light with background light like PMTs would.  G-APDs have already been used to detect Cherenkov photons using MAGIC \citep{Biland07}, and the dedicated First G-APD Cherenkov Telescope (FACT) uses G-APDs \citep{Braun09,Anderhub11}.  

A developing possibility is to use new cryogenic photon counting detector technologies, such as superconducting tunnel junction (STJ) detectors and Transition Edge Sensors (TES) \citep[e.g.,][]{Verhoeve08}.  STJ detectors have small pixel sizes compared to PMTs ($\sim 30\ \mu m$), high quantum efficiency, extremely high time resolution ($\sim 1\ \mu s$), and even some low spectral resolution.  On the other hand, the detectors must be cooled to sub-Kelvin temperatures.  S-Cam at the William Herschel Telescope has demonstrated that STJs can be used for high time resolution astronomy \citep[e.g.,][]{Perryman99,Oosterbroek06}; however, it only consists of a few pixels and has a very small field of view \citep[e.g.,][]{Verhoeve06}.  Much larger arrays are needed to fit the PSF of even one source imaged with an IACT, and differential photometry with STJs at IACTs is vastly more difficult still.  Finally, STJs become saturated at count rates of $\sim 10^3~\sec^{-1}$ per pixel.  Assuming a pixel size of $30\ \mu m$, there are $n_{\rm pix} \approx 2 \times 10^5 (\theta_{\rm PSF} / 2\ \arcmin)^2$ pixels in the image of a source, and from eq.~\ref{eqn:NphStar}, we see saturation occurs at $V \approx 8.3$.  Thus, observations with current STJ technologies might require going to fainter sources on moonless nights.  High time resolution photometry (and spectroscopy) has also been demonstrated with TES studies of the Crab pulsar, but again only for small detectors \citep[e.g.,][]{Romani99}. 

\subsection{Systematics and Challenges}
\label{sec:Systematics}
In practice, a number of systematic issues present the true challenge to attaining sub-millimagnitude photometry.  These systematic effects add time-correlated red noise to the photometry, and decrease the precision.  

Changes in the PSF as the telescope tracks a target may be one source of systematic noise.  \citet{Gillon09} attempted to use the defocused VLT to measure the light curves of WASP-4 and WASP-5.  They found that systematic trends in the light curves of the target stars and field stars, probably due to the active optics being turned off.  IACTs may experience similar effects; the PSFs of Whipple and HESS are changed by distortions in its support structure \citep{Cornils03,Kildea07}, although this seems to be insignificant for CANGAROO-II \citep{Kawachi01}.  To evade such problems, differential photometry with other stars in the field is probably necessary.

Photon-limited millimagnitude differential photometry is regularly achieved on the ground, but this requires monitoring multiple stars in the field \citep{Everett01,Hartman05}.  However, at $V = 8$, the population of stars is relatively sparse ($\sim 0.3 - 2\ {\rm deg}^{-2}$ at $|b| \ge 20^{\circ}$; \citealt{Bahcall80}), with only a few in the entire $\sim 5^{\circ}$ field of view (which would have to be covered with detectors).  The situation improves somewhat at greater magnitudes (at $V = 10$, $\sim 2 - 10\ {\rm deg}^{-2}$ spacing implies $\sim 20 - 50\ \arcmin$ between stars; at $V = 12$, $\sim 10 - 70\ {\rm deg}^{-2}$ implies $8 - 20\ \arcmin$ spacing; \citealt{Bahcall80}).  Furthermore, the PSF of IACTs typically grows larger further from the centre of the field, which may hamper differential photometry.  Image subtraction techniques have been invented for photometry when there are PSF variations in a field, however \citep{Alard00,Kerins10}.  A wide-field ($15^{\circ}$) IACT using Schmidt optics has been proposed, which would have a stable PSF over its entire field; this kind of IACT may prove useful for differential photometry \citep{Mirzoyan09}.  

Another potential problem is that the comparison stars will appear on different CCDs than the target source.  Since there will be gaps between the detectors (whether CCDs, PMTs, G-APDs, or SJDs), the camera may have to be rotated during the night to avoid the comparison stars' images from sliding into the gaps.  Otherwise, as the fraction of a reference star's PSF in a gap changes, there will be apparent changes in its brightness.  On the other hand, Cherenkov telescopes typically use light concentrators like Winston cones to funnel light into detectors with little dead space, including HESS \citep{Bernlohr03}, MAGIC \citep{Lucarelli08}, and VERITAS \citep{Weekes02}, and potentially, the future CTA \citep{CTA10}.  Finally, the guiding will have to be sufficiently accurate that the image does not wander significantly during an observation.  

Stray light in the telescope and detector can also introduce systematic errors in fine photometry (Deil 2011, private communication).  For differential photometry, we are concerned with temporal and spatial \emph{variations} of stray light, since we are interested in relative changes of the brightness of the source and stray light introduces much less shot noise than the night sky background itself.  These variations can be caused by either (1) the amount of stray light varying as the telescope tracks a source over the course of a night and (2) different amounts of stray light between the comparison star and the source because they are at different angles.  Winston cones, which reduces the solid angle visible to the detectors, can reduce the amount of light from the sky or the ground around the telescope; in HESS, they accept light roughly from the angle subtended by the primary reflector (acceptance angle is $27^{\circ}$), although small amounts of light come from larger angles \citep{Bernlohr03}.  In addition, there are gaps between the mirror facets in the reflector, through which the mount is visible.  This is partly ameliorated in HESS because the mount is painted red while PMTs are mostly sensitive to blue light \citep{Bernlohr03}.  However, \citet{Preu02} find that ground light is about $\sim 15\%$ of the night sky background for HESS, with the telescope mount contributing an additional $\sim 3\%$.  Thus, in order to ensure that stray light is less than $10^{-3}$ of the source light, the source may have to be $\sim 200$ times brighter than the night sky, or $V \la V_{\rm sky} - 5.7 \approx 2.7 - 4.7$.  On the other hand, there are techniques for doing photometry in the presence of variable backgrounds included stray light \citep[e.g.,][]{Kerins10}, and the variations in the stray light over the course of an observation may a small fraction of the total amount of stray light, allowing photometry of fainter objects.

These systematics introduce red noise that is correlated on a time scale of a few hours, and make detections harder even for periodic phenomena.  \citet{Pont06} considered the effects of red noise specifically in the case of exoplanet transits, although their analysis should work similarly for other periodic phenomena.  If a transit is observed $N_{\rm tr}$ times and has a depth $d$, the threshold for detection is
\begin{equation}
d^2 \ga \frac{S^2 \sum_{k = 1}^{N_{\rm tr}} n_k^2 (\sigma_w^2 / n_k + \sigma_r^2)}{n^2},
\end{equation}
where $S \approx 10$ is a typical signal-to-noise required for detection, $n_k$ is the number of observations within a transit, $n = \sum_{k = 1}^{N_{\rm tr}} n_k \approx N_{\rm tr} n_k$ is the total number of observations in all the transits, $\sigma_w$ is the time-uncorrelated white noise (including photon, scintillation, and detector noise), and $\sigma_r$ is the red noise.  When white noise dominates ($\sigma_w \ga \sqrt{n_k} \sigma_r$), $d \ga S \sigma_w / \sqrt{n}$, but when red noise dominates, $d \ga S \sigma_r / \sqrt{N_{\rm tr}}$.  For most transit surveys, $\sigma_r \approx 0.001$, and since IACTs reach $\sigma_w \approx 0.001$ within seconds (eq.~\ref{eqn:sigmaVPoissonBright}--\ref{eqn:sigmaVScint}), they may be in the red noise regime, where only the number of transits observed matters:
\begin{equation}
d \ga 0.01 \left(\frac{\sigma_r}{0.001}\right) \left(\frac{S}{10}\right) \frac{1}{\sqrt{N_{\rm tr}}}.
\end{equation}
This negates most of the advantages of IACTs over smaller telescopes, but the remaining advantages from the shorter exposure times would be: (1) IACTs could do timing studies more accurately; (2) IACTs could move to the next target more quickly (though subject to slew times); (3) IACTs could easily conduct multiband photometry.  Furthermore, photometry of individual transiting exoplanets have been able to reduce the red noise to $\sim 10^{-4}$ (e.g., \citealt*{Winn07}; \citealt{Winn09}).  Note also that most IACTs come in arrays of multiple telescopes.  It is unknown if the red noise between telescopes in an array is correlated, but if it is not, each telescope would contribute an independent observation, increasing $N_{\rm tr}$ by the number of IACTs.  For CTA, which is planned to have $\sim 50$ IACTs, this effect may be very powerful.  

Finally, absolute photometry over weeks or longer requires the monitoring the reflectance of the IACTs.  IACTs are built in the open, exposed to the elements, and suffer decreased reflectivity with time.  IACTs are therefore occasionally washed or re-aluminised, which will affect performance \citep{Kildea07}.  

\section{Spectroscopy with IACTs}
\label{sec:Spectroscopy}
As light buckets, IACTs could also be useful for spectroscopy.  If a proper spectrometer could be built, they can rapidly generate high $S/N$ spectra:
\begin{equation}
S/N \approx \left\{ \begin{array}{l}
		\displaystyle 416 \times 10^{-0.2 (V - 8)} \sqrt{\frac{A_{100} t_{\rm int} \eta_{0.5}}{{\cal R}_{10000}}} \\
\hfill(V \la V_{\rm sky})\\
\\
		\displaystyle 491 \times 10^{-0.2 (2V - 16 - V_{\rm sky} + 8.36)} \sqrt{\frac{A_{100} t_{\rm int} \eta_{0.5}}{{\cal R}_{10000}}}\\
\hfill(V \ga V_{\rm sky}),
		\end{array} \right.
\end{equation}
where ${\cal R} = 10^4 {\cal R}_{10000} = \Delta\lambda / \lambda$ is the spectral resolution.  Measurements of spectral line structure and depth are differential, thus there may not be as many systematics from the atmosphere.  

\begin{figure}
\centerline{\includegraphics[width=8cm]{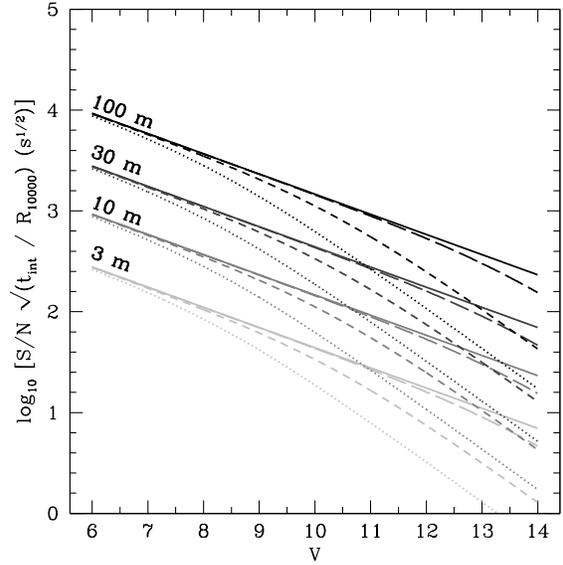}}
\caption{Signal-to-noise achieved during spectroscopy by telescopes with different apertures and PSF sizes, as a function of $V$.  The line styles are the same as in Fig.~\ref{fig:sigmaV}.\label{fig:SNSpectra}}
\end{figure}

The main challenge to designing a dispersive spectrograph for an IACT is that the low angular resolution smears out the image of the source.  The resolving power of a spectroscopic grating is ${\cal R} \approx k \lambda d_1 / (\sigma \theta_{\rm slit} D_T)$, where $d_1$ is the size of the beam on the grating, $\sigma$ is the grating spacing, $\theta_{\rm slit}$ is the slit width in terms of sky angular diameter, $D_T$ is the telescope aperture, and $k \approx 1$ is a combination of other parameters relating to the optics of the spectrograph \citep[e.g.,][]{Schroeder00}.  For typical parameters,
\begin{equation}
{\cal R} ({\rm grating}) \approx  47 k \left(\frac{d_1}{1\ \meter}\right) \left(\frac{\sigma^{-1}}{1000\ \mm^{-1}}\right) \left(\frac{\theta_{\rm slit}}{4\ \arcmin}\right)^{-1} \left(\frac{D_T}{10\ \meter}\right)^{-1}.
\end{equation}
Echelle spectrographs can do somewhat better by using higher orders, with a spectral resolution ${\cal R} \approx 2 d_1 \tan \delta / (\theta_{\rm slit} D_T)$, where $\delta$ is the groove angle \citep[e.g.,][]{Schroeder00}.  We find, with typical $\delta \approx 63.5^{\circ}$
\begin{equation}
\label{eqn:REchelle}
{\cal R} ({\rm echelle}) \approx 340 \left(\frac{d_1}{1\ \meter}\right) \left(\frac{\theta_{\rm slit}}{4\ \arcmin}\right)^{-1} \left(\frac{D_T}{10\ \meter}\right)^{-1}.
\end{equation}
Reducing the slit angular size below $2 \theta_{\rm PSF}$ sacrifices light from the target object, which negates the advantage of the large apertures of IACTs.  In principle, this could be evaded by integral field spectroscopy, with the source image divided into many small pieces which are fed through narrow slits with fibre optics.  This would require a large number of fibres to achieve high resolution: for the echelle case, $N_{\rm fib} \approx 870 ({\cal R} / 10^4)^2 (D_T / 10\ \meter)^2 (\theta_{\rm PSF} / 2\ \arcmin)^{2}$.  Even higher ${\cal R}$ may be possible: VIRUS on the Hobby-Eberly Telescope demonstrates integral field spectroscopy with $\sim 30000$ fibres \citep{Hill06}.  However, such instruments may actually be more complicated and expensive than the Cherenkov telescope itself, and might require a dedicated IACT-like telescope.  Given the extreme costs of ELTs and the applications of a high ${\cal R}$ spectrograph for even just bright sources, however, the possibility of massive integral field spectroscopy applied to source images in IACT-like telescopes warrants further investigation.

Finally, new cryogenic spectrophotometer technologies like STJs and TESs may also be useful for high time resolution spectroscopy at low ${\cal R}$ (up to $\sim 100$ for TESs).  They have already demonstrated the capability of detecting some spectral lines \citep[e.g.,][]{deBruijne02,Reynolds05} and measuring time variations in spectra in pulsars \citep{Romani01}.

\section{Polarimetry with IACTs}
\label{sec:Polarimetry}
Polarimetry has now reached the $10^{-6}$ level with instruments like PlanetPol \citep{Hough06} and POLISH \citep{Wiktorowicz08}.  Polarimetry is typically limited by the number of photons, which IACTs excel at maximizing for bright objects. 

Polarimeters work by measuring the changes in intensity as the polarization angle of transmitted light is varied (for example, by rotating a polarizer).  As a model of a linear polarimeter, suppose it measures the intensity of a source parallel and perpendicular to its angle of polarization over an integration time $t_{\rm obs}$.  In practice, since we do not know the direction of linear polarization of a source (that is, we have to fit two Stokes parameters), we have to observe it at two polarizations tilted $45^\circ$ from parallel, so that the total number of photons collected is $N = N_{\|} + N_{\bot} + N_{\nearrow} + N_{\nwarrow}$.  The total number of photons observed for each polarization $i \in \{\|, \bot, \nearrow, \nwarrow\}$ are $N_{\rm obs}^i = N_{\star}^i + N_{\rm back}^i$ where $N_{\star}^i$ is the number of photons from the source and $N_{\rm back}^i$ is the number of photons from the (sky) background.  Since the sky polarization is typically large ($\sim 50$ per cent), it must be subtracted off by taking another measurement of the sky itself.  The polarimeter observes the sky background for some time $t_{\rm sky}$, measuring $N_{\rm sky}^{\|}$ and $N_{\rm sky}^{\bot}$ photons from the sky; the background is then estimated as $N_{\rm back}^i = (t_{\rm obs} / t_{\rm sky}) N_{\rm sky}^i$.  After sky subtraction, the degree of polarization is then $p_{\star} = (N_{\star}^{\|} - N_{\star}^{\bot}) / (N_{\star}^{\|} + N_{\star}^{\bot})$, or
\begin{equation}
p_{\star} = \displaystyle \frac{[N_{\rm obs}^{\|} - \frac{t_{\rm obs}}{t_{\rm sky}} N_{\rm sky}^{\|}] - [N_{\rm obs}^{\bot} - \frac{t_{\rm obs}}{t_{\rm sky}} N_{\rm sky}^{\bot}]}{[N_{\rm obs}^{\|} - \frac{t_{\rm obs}}{t_{\rm sky}} N_{\rm sky}^{\|}] + [N_{\rm obs}^{\bot} - \frac{t_{\rm obs}}{t_{\rm sky}} N_{\rm sky}^{\bot}]}.
\end{equation}
The Poisson noise in each of the measured photon numbers contributes to the noise of the measured $p_{\star}$: $\sigma_{p_{\star}}^2 = \sum_{j \in \{\|, \bot\}} \left(\frac{\partial p_{\star}}{\partial N_{\rm obs}^j} \sigma_{N_{\rm obs}^j}\right)^2 + \left(\frac{\partial p_{\star}}{\partial N_{\rm sky}^j} \sigma_{N_{\rm sky}^j}\right)^2$.  This gives us:
\begin{equation}
\sigma_{p_{\star}} = \sqrt{\sum_{j \in \{\|, \bot\}} 4 \frac{(N_{\star}^{\bar{j}})^2}{(N_{\star}^{\|} + N_{\star}^{\bot})^4} \left[N_{\rm obs}^j + \left(\frac{t_{\rm obs}}{t_{\rm sky}}\right)^2 N_{\rm sky}^j\right]},
\end{equation}
defining $\bar{j}$ as $\bot$ when $j = \|$ and $\|$ when $j = \bot$.  When both the sky and source are weakly polarized ($N \approx 4 N^{\bot} \approx 4N^{\|} \approx 4N^{\nearrow} \approx 4N^{\nwarrow}$), $\sigma_{p_{\star}} \approx \sqrt{2[N_{\rm obs} + (t_{\rm obs}/t_{\rm sky})^2 N_{\rm sky}] / N_{\star}^2}$.  The noise can be understood to mainly arise from either sky subtraction or the observation itself:
\begin{equation}
\label{eqn:sigmaPUnevaluated}
\sigma_{p_{\star}} \approx \left\{ \begin{array}{ll} 
\displaystyle \frac{t_{\rm obs} \sqrt{2N_{\rm sky}}}{t_{\rm sky} N_{\star}} & (t_{\rm sky} \la t_{\rm obs} / \sqrt{1 + N_{\star} / N_{\rm sky}})\\
\displaystyle \frac{\sqrt{2N_{\rm obs}}}{N_{\star}} & (t_{\rm sky} \ga t_{\rm obs} / \sqrt{1 + N_{\star} / N_{\rm sky}})
\end{array} \right. .
\end{equation}
Essentially, when the sky subtraction measurement is long enough, the degree of polarization of the combined source and sky can be measured to an accuracy $\sqrt{2/N_{\rm obs}}$.  If the source dominates, this is the accuracy of the source polarization measurement; if the sky dominates then the source polarization is diluted by a factor $N_{\rm sky} / N_{\star}$ and the accuracy in the measurement is reduced.  While measurements of optical circular polarization are relatively rare, they would be conceptually similar: we would simply take the difference of left-handed and right-handed photons, and with only one Stokes parameter, the noise would be $\sqrt{2}$ times smaller for the same total observation time.  

Using equation~\ref{eqn:NphStar}, we can now evaluate how accurately IACTs can measure polarization, based on the photon noise and assuming the sky subtraction is sufficiently long ($t_{\rm sky} \ga t_{\rm obs} / \sqrt{1 + N_{\star}/N_{\rm sky}}$):
\begin{equation}
\label{eqn:sigmaPNumerical}
\sigma_{p_{\star}} \approx \left\{ \begin{array}{l}
8.7 \times 10^{-5} \times 10^{0.2(V - 8)} / \sqrt{A_{100} t_{\rm int} \eta_{0.5}}\\
\hfill(V \la V_{\rm sky})\\
\\
7.3 \times 10^{-5} \times 10^{-0.2(V_{\rm sky} - 8.36 - 2V + 16)} / \sqrt{A_{100} t_{\rm int} \eta_{0.5}}\\
\hfill(V \ga V_{\rm sky})
\end{array} \right.
\end{equation}
We see the photon statistics are sufficient for a single IACT to reach $10^{-6}$ precision in one hour exposures for a $V = 8$ target.  

The angle of linear polarization $\theta$ can also be measured, using the formula $2\theta = \tan^{-1} (u/q)$, where $u$ and $q$ are the normalised Stokes parameters.  The telescope can estimate these parameters as $u = (N_{\star}^{\nearrow} - N_{\star}^{\nwarrow}) / (N_{\star}^{\nearrow} + N_{\star}^{\nwarrow})$ and $q = (N_{\star}^{\|} - N_{\star}^{\bot}) / (N_{\star}^{\|} + N_{\star}^{\bot})$.  It can then be shown that when the sky and source polarization are weak, the photon noise introduces an uncertainty in the angle equal to the values of $\sigma_{p_{\star}}$ in equations~\ref{eqn:sigmaPUnevaluated} divided by $2 p_{\star}$.  If the sky exposure is long enough ($t_{\rm sky} \ga t_{\rm obs} / \sqrt{1 + N_{\star} / N_{\rm sky}}$), the uncertainty in the measured polarization angle due to Poisson noise is:
\begin{equation}
\label{eqn:sigmaThetaNumerical}
\sigma_{\theta} \approx \left\{ \begin{array}{l}
0.25^{\circ} \times 10^{0.2(V - 8)} / \sqrt{A_{100} t_{\rm int} \eta_{0.5} p_{0.01}}\\
\hfill(V \la V_{\rm sky})\\
\\
0.21^{\circ} \times 10^{-0.2(V_{\rm sky} - 8.36 - 2V + 16)} / \sqrt{A_{100} t_{\rm int} \eta_{0.5} p_{0.01}}\\
\hfill(V \ga V_{\rm sky}),
\end{array} \right.
\end{equation}
where $p_{0.01} = p_{\star} / 0.01$.

As with photometry, the systematics are unknown and may dominate.  Because the sky is highly polarized, it will have to be subtracted to a very high precision; for comparison, \citet{Huovelin90} demonstrated the ability to subtract off the sky to $1/500$ precision.  \citet{Hough06} found that sky noise was negligible with 5 arcsec apertures except on moonlit nights, but an IACT would effectively have $\sim 600$ times greater sky noise.  More severely, the degree and direction of polarization change with direction on the sky; this effect on sky channel observations, which would have to be at least a few arcminutes away for an IACT, would introduce some systematic errors.  \citet{Fox92} estimate (with a sky observation 4 arcmin away and using 15 arcsec apertures) that this effect could be severe on moonlit nights for stars with $V \ga 3$, although it could be much smaller on moonless nights.  Furthermore, polarization induced by the Cherenkov telescopes themselves may limit polarimetry.  Aluminium-coated mirrors with revolution symmetry induce a polarization signal of $\sim 3 \times 10^{-8} (\delta\theta / 1\ \arcmin)^2$, where $\delta\theta$ is the angle between the revolution axis and the source, if the integrated light from the entire PSF is considered \citep{SanchezAlmeida92}.  The individual facets of IACTs are tilted at angles of up to $\sim 10^{\circ}$ degrees \citep[e.g.,][]{Bernlohr03}, so they individually induce polarizations of order $\sim 10^{-4} - 10^{-2}$.  However, the overall shape of an IACT primary reflector is parabolic; if the polarizations of the facets cancel each other out, the remaining polarization could be very small ($10^{-8} - 10^{-7}$).  These systematics need to be considered further before use of IACTs as high precision optical polarimeters.  In any case, high-speed lower-precision polarimetry of bright transients like gamma-ray bursts (GRBs), seems possible with IACTs. \footnote{HEGRA previously measured the polarization of Cherenkov pulses to within a few percent \citep{Doering01}.}

\section{Applications}
The low angular resolution of IACTs generally makes them suitable for bright sources or small integration times, in contrast to typical ELTs which can go down to very faint magnitudes. Although this makes IACTs unsuitable for much ELT science, there are still several interesting targets of sufficient brightness that might be studied by IACTs.  These span a range of different science opportunities, given below:

\emph{Occultations} -- Lunar occultations of bright stars are commonly used to measure stellar diameters \citep[e.g.,][]{Richichi94}.  These measurements do not require difficult millimagnitude differential photometry, but instead rapid measurements of large magnitude changes.  Indeed, \citet{Deil09} has already shown that IACTs excel for such situations.

A more exotic possibility is occultations of bright stars by kilometre-scale Kuiper Belt Objects.  Diffraction around these objects produces variability at the several percent level (e.g., \citealt*{Roques87}; \citealt{Bianco09,Schlichting09}).  The brightness fluctuation is of order a few percent and lasts a fraction of a second \citep[e.g.,][]{Roques00}.  An array of telescopes, as in most IACTs, could probe the two dimensional structure of the diffraction pattern.  In the Fresnel regime, this provides an independent measure of the occulting body's distance; combined with the transit time, the velocity is then determined.

\emph{Exoplanets} -- Exoplanet transits occur when a planet occults a star from the vantage point of Earth, causing its apparent brightness to drop a small amount (1\% or less for sun-like stars).  They are essentially periodic, occuring once per planet orbit and last for a few hours.  IACTs have \emph{already} conducted $\sim 1$ per cent optical photometry with the Crab pulsar \citep{Hinton06,Lucarelli08}, thus demonstrating in principle they can observe transits of Jupiter-sized exoplanets around sun-like stars.  However, the systematics are probably different for the $\sim$ hour long observations of exoplanet transits and the sub-second Crab pulsations.  IACTs are not well-suited for transit searches \citep*[at which small telescopes perform better; see][]{Pepper03}, but instead for follow-up.  With their short integration times, IACTs can finely sample transit light curves for Transit Timing Variation and Transit Duration Variation studies \citep{MiraldaEscude02,Holman05,Agol05}.  These variations are expected to be of order seconds to minutes, compared to the hour time-scales of the transit itself.  If the technical challenges of differential photometry (\S~\ref{sec:Systematics}) can be overcome and photon-limited statistics achieved, the transits of planets around giant stars \citep*[e.g.,][]{Assef09} or small planets around sun-like stars might be detectable.  

Transiting exoplanets can also be studied spectroscopically, including detecting and characterising the atmospheres of transiting exoplanets \citep{Brown01,Charbonneau02}.  However, these applications generally require ${\cal R} \approx 10^4$ and therefore a multi-fibre instrument to harness the light collecting power of an IACT.

Finally, IACTs could quickly collect the photons needed for detecting the polarized reflected light of an exoplanet \citep*{Seager00} or the changes in a star's polarization during an exoplanet transit \citep{Carciofi05}.  A claimed polarimetric detection of HD 189733b is controversial \citep{Berdyugina08,Wiktorowicz09}, but, if real, the detection appears to be strongest in blue bands because of Rayleigh scattering in the exoplanet's atmosphere \citep{Berdyugina11}, and IACTs have an advantage at bluer bands where the dark sky backgrounds are lower.

\emph{Stars} -- The low angular resolution of IACTs limits them to the brightest and nearest stars, but IACTs remain suitable for monitoring their short time-scale photometric variability.  The brightest stellar flares reach $\Delta V \sim 0.1$ and last for minutes to days (\citealt*{Schaefer00}; \citealt{Bentley09}).  IACTs are potentially useful for binary star studies; photon-limited photometry could even be used to detect relativistic beaming effects (e.g., \citealt{Loeb03}; \citealt*{Zucker07}).  

Most nearby stars have very small intrinsic polarizations ($\la 10^{-5}$), but interstellar (e.g., \citealt{Tinbergen82}; \citealt*{Bailey10}) dust can be studied through induced polarizations.  The brightest X-ray binaries, like Cygnus X-1, have $V \approx 9$, and might be studied polarimetrically by IACTs to constrain the inclination of the binary orbit \citep[c.f.][]{Wiktorowicz08}.

\emph{Gamma-ray bursts} -- The brightest GRBs approach or even exceed $V_{\rm sky}$ for time periods of minutes \citep[such as GRB080319B, peaking at $V \approx 5.3$;][]{Racusin08,Bloom09}.  MAGIC has a fast slew time of $\la$30 seconds for anywhere on the sky \citep{Bretz03}, while HESS and VERITAS have slew times of 2-3 minutes for any location on the sky \citep{Hinton04,Humensky07}.  With their huge collection areas, IACTs could construct millimagnitude optical light curves of GRBs or other transients within the first few minutes of their triggers, with 100 millisecond sampling times.  As has been suggested before \citep[e.g.,][]{Beskin99,OnaWilhelmi04,Hinton06}, such observations could even be concurrent with TeV observations \citep[e.g.,][]{Aharonian09}, if the central pixel is used for optical photometry and the remaining field of view for gamma-ray observations, as with MAGIC \citep{Lucarelli08}.  

Spectroscopy of GRBs with IACTs could rapidly constrain the redshifts and properties of gas in the host galaxy and intervening intergalactic medium.  This may prove useful in time-dependent studies of the geometry of intergalactic metal-line absorbers on GRB sightlines \citep[e.g.,][]{Frank07,DElia10}.

Finally, the optical polarization of GRB afterglows is a powerful discriminant of emission mechanisms and the magnetic field structure in GRB afterglows \citep[e.g.,][]{Gruzinov99}.  Observations of afterglows on time-scales of hours to days typically find polarizations of order a few percent \citep[e.g.,][]{Covino99}, with some afterglows reaching $\sim 10$ per cent linear polarization \citep[e.g.,][]{Bersier03}.  However, there are few constraints on the early optical polarization of GRB afterglows; \citet{Mundell07} found an upper limit of $\sim 8$ per cent on optical polarization a few minutes after GRB 060418.  IACTs can rapidly collect the photons needed for early time polarization measurements for the brightest GRBs to relatively high precisions (from eq.~\ref{eqn:sigmaPNumerical}, $\la 1$ per cent in 1 second, even for $V \approx 13$ on moonlit nights, $V \approx 15$ on moonlit nights). 

\section{Conclusions}
Diffraction-limited ELTs have many advantages, in that they can examine faint sources necessary for studies involving areas like cosmology, distant Galactic stars, and direct imaging of exoplanets, but the telescope itself is very costly.  IACTs already have achieved ELT-scale apertures and are very inexpensive, but are only effective for bright sources, and the low angular resolution means that complex instruments are necessary.  For high spectral resolution optical spectroscopy in particular, the instrument may in fact be more expensive than the IACT itself.  Between the two extremes, telescopes of intermediate optical quality may use simpler instrumentation while still being relatively cheap.  A telescope with $\theta_{\rm PSF} = 10\ \arcsec$, for example, would have no disadvantage down to $V = 14$ even on moonlit nights.  The smaller source images may make techniques for high speed CCD imaging more practical \citep{ODonoghue95}.  Increased angular resolution would also facilitate higher ${\cal R}$ spectroscopy, up to $\sim 10^4$ with an echelle and a single slit (eq.~\ref{eqn:REchelle}).  Finally, polarimeters like PlanetPol \emph{already} have demonstrated that precise polarimetry can be done with $\sim 5\ \arcsec$ apertures, even on moonlit nights \citep{Hough06}.  The downside of increased angular resolution would be that bright sources would saturate more quickly.

An IACT-like telescope in space would have many advantages. All scintillation and systematic noise due to the atmosphere would vanish.  A spaceborne IACT-like telescope could therefore take full advantage of the massive light collecting power to approach \emph{micromagnitude} photometry of $V = 8$ sources within a few hours of integration time.  Alternatively, crude ($\sigma_V \approx 0.1$) \emph{microsecond} photometry could be achieved \citep[c.f.][]{Deil09}.  The sky background for polarimetry would likewise mostly vanish.  \emph{Herschel} demonstrates that $\sim 4$ metre telescopes are viable in space, and optical telescopes would need sophisticated cooling systems only for the detector.  Furthermore, spaceborne telescopes would evade NIR skyglow \citep{Leinert98}, which might prove useful in investigating secondary eclipses of exoplanets \citep[e.g.,][]{Charbonneau05}.  

\section*{Acknowledgments}
I am especially grateful to Scott Gaudi for encouragement and discussions.  I also thank Christoph Deil, Andy Gould, David Hanna, Christopher Kochanek, Fabrizio Lucarelli, Stephan LeBohec, Paul Martini, Rene Ong, Kris Stanek, and Todd Thompson  for additional useful comments and discussion.  This work is funded in part by an Elizabeth Clay Howald Presidential Fellowship from OSU.


\begin{thebibliography}{}

\bibitem[\protect\citeauthoryear{Agol et al.}{2005}]{Agol05} Agol E., Steffen J., Sari R., \& Clarkson W.\ 2005, \mnras, 359, 567 

\bibitem[\protect\citeauthoryear{Aharonian \& Akerlof}{1997}]{Aharonian97} Aharonian, F.~A., \& Akerlof, C.~W.\ 1997, Annual Review of Nuclear and Particle Science, 47, 273 

\bibitem[\protect\citeauthoryear{Aharonian et al.}{2008}]{Aharonian08} Aharonian, F., Buckley, J., Kifune, T., \& Sinnis, G.\ 2008, Reports on Progress in Physics, 71, 096901 

\bibitem[\protect\citeauthoryear{Aharonian et al.}{2009}]{Aharonian09} Aharonian F., et al.\ 2009, \aap, 495, 505 

\bibitem[\protect\citeauthoryear{Akhperjanian et al.}{1998}]{Akhperjanian98} Akhperjanian A., Kankanian R., Sahakian V., Heusler A., Wiedner C.-A., \& Wirth H.\ 1998, Experimental Astronomy, 8, 135 

\bibitem[\protect\citeauthoryear{Alard}{2000}]{Alard00} Alard C.\ 2000, \aaps, 144, 363 

\bibitem[\protect\citeauthoryear{Anderhub et al.}{2009}]{Anderhub09} Anderhub, H., et al.\ 2009, Journal of Instrumentation, 4, 10010 

\bibitem[\protect\citeauthoryear{Anderhub et al.}{2011}]{Anderhub11} Anderhub, H., et al.\ 2011, Nuclear Instruments and Methods in Physics Research A, 639, 58 

\bibitem[\protect\citeauthoryear{Assef, Gaudi, \& Stanek}{Assef et al.}{2009}]{Assef09} Assef R.~J., Gaudi B.~S., \& Stanek K.~Z.\ 2009, \apj, 701, 1616 

\bibitem[\protect\citeauthoryear{Bahcall \& Soneira}{1980}]{Bahcall80} Bahcall J.~N., \& Soneira R.~M.\ 1980, \apjs, 44, 73 

\bibitem[\protect\citeauthoryear{Bailey, Lucas, \& Hough}{Bailey et al.}{2010}]{Bailey10} Bailey J., Lucas P.~W., \& Hough J.~H.\ 2010, \mnras, 405, 2570 

\bibitem[\protect\citeauthoryear{Bentley et al.}{2009}]{Bentley09} Bentley S.~J., Hellier C., Maxted P.~F.~L., Dhillon V.~S., Marsh T.~R., Copperwheat C.~M., \& Littlefair S.~P.\ 2009, \aap, 505, 901 

\bibitem[\protect\citeauthoryear{Berdyugina et al.}{2008}]{Berdyugina08} Berdyugina S.~V., Berdyugin A.~V., Fluri D.~M., \& Piirola V.\ 2008, \apj, 673, L83 

\bibitem[\protect\citeauthoryear{Berdyugina et al.}{2011}]{Berdyugina11} Berdyugina S.~V., Berdyugin A.~V., Fluri D.~M., \& Piirola V.\ 2011, \apj, 728, L6 

\bibitem[\protect\citeauthoryear{Bernl{\"o}hr et al.}{2003}]{Bernlohr03} Bernl{\"o}hr K., et al.\ 2003, Astroparticle Physics, 20, 111 

\bibitem[\protect\citeauthoryear{Bersier et al.}{2003}]{Bersier03} Bersier D., et al.\ 2003, \apj, 583, L63 

\bibitem[\protect\citeauthoryear{Beskin et al.}{1999}]{Beskin99} Beskin G.~M., et al.\ 1999, \aaps, 138, 589 

\bibitem[\protect\citeauthoryear{Bessell}{2005}]{Bessell05} Bessell M.~S.\ 2005, \araa, 43, 293 

\bibitem[\protect\citeauthoryear{Bianco et al.}{2009}]{Bianco09} Bianco F.~B., Protopapas P., McLeod B.~A., Alcock C.~R., Holman M.~J., \& Lehner M.~J.\ 2009, \aj, 138, 568 

\bibitem[\protect\citeauthoryear{Biland et al.}{2007}]{Biland07} Biland, A., et al.\ 2007, Nuclear Instruments and Methods in Physics Research A, 581, 143 

\bibitem[\protect\citeauthoryear{Bloom et al.}{2009}]{Bloom09} Bloom J.~S., et al.\ 2009, \apj, 691, 723 

\bibitem[\protect\citeauthoryear{Borra}{2010}]{Borra10} Borra E.~F.\ 2010, \apj, 715, 589 

\bibitem[\protect\citeauthoryear{Braun et al.}{2009}]{Braun09} Braun, I., et al.\ 2009, 
Nuclear Instruments and Methods in Physics Research A, 610, 400 

\bibitem[\protect\citeauthoryear{Bretz et al.}{2003}]{Bretz03} Bretz, T., Dorner, D., Wagner, R., \& MAGIC collaboration 2003, International Cosmic Ray Conference, 5, 2943 

\bibitem[\protect\citeauthoryear{Brown}{2001}]{Brown01} Brown T.~M.\ 2001, \apj, 553, 1006 

\bibitem[\protect\citeauthoryear{Buzhan et al.}{2003}]{Buzhan03} Buzhan, P., et al.\ 2003, Nuclear Instruments and Methods in Physics Research A, 504, 48 

\bibitem[\protect\citeauthoryear{Carciofi \& Magalh{\~a}es}{2005}]{Carciofi05} Carciofi A.~C., \& Magalh{\~a}es A.~M.\ 2005, \apj, 635, 570 

\bibitem[\protect\citeauthoryear{Charbonneau et al.}{2002}]{Charbonneau02} Charbonneau D., Brown T.~M., Noyes R.~W., \& Gilliland R.~L.\ 2002, \apj, 568, 377 

\bibitem[\protect\citeauthoryear{Charbonneau et al.}{2005}]{Charbonneau05} Charbonneau D., et al.\ 2005, \apj, 626, 523 

\bibitem[\protect\citeauthoryear{Colavita, Shao, \& Staelin}{Colavita et al.}{1987}]{Colavita87} Colavita M.~M., Shao M., \& Staelin D.~H.\ 1987, Applied Optics, 26, 4106 

\bibitem[\protect\citeauthoryear{Cornils et al.}{2003}]{Cornils03} Cornils R., et al.\ 2003, Astroparticle Physics, 20, 129 

\bibitem[\protect\citeauthoryear{Covino et al.}{1999}]{Covino99} Covino S., et al.\ 1999, \aap, 348, L1 

\bibitem[\protect\citeauthoryear{CTA Consortium}{2010}]{CTA10} CTA Consortium, 2010, arXiv:1008.3703 

\bibitem[\protect\citeauthoryear{de Bruijne et al.}{2002}]{deBruijne02} de Bruijne J.~H.~J., et al.\ 2002, \aap, 381, L57 

\bibitem[\protect\citeauthoryear{Deil et al.}{2009}]{Deil09} Deil C., Domainko W., Hermann G., Clapson A.~C., F{\"o}rster A., van Eldik C., \& Hofmann W.\ 2009, Astroparticle Physics, 31, 156 

\bibitem[\protect\citeauthoryear{D'Elia et al.}{2010}]{DElia10} D'Elia V., et al.\ 2010, \mnras, 401, 385 

\bibitem[\protect\citeauthoryear{Dhillon et al.}{2007}]{Dhillon07} Dhillon V.~S., et al.\ 2007, \mnras, 378, 825 

\bibitem[\protect\citeauthoryear{Doering et al.}{2001}]{Doering01} Doering M., Bernloehr K., Hermann G., Hofmann W., \& Lampeitl H.\ 2001, arXiv:astro-ph/0107149 

\bibitem[\protect\citeauthoryear{Doro \& CTA consortium}{2009}]{Doro09} Doro M., \& CTA consortium, f.~t.\ 2009, arXiv:0908.1410 

\bibitem[\protect\citeauthoryear{Dravins et al.}{1997a}]{Dravins97} Dravins D., Lindegren L., Mezey E., \& Young A.~T.\ 1997a, \pasp, 109, 173 

\bibitem[\protect\citeauthoryear{Dravins et al.}{1997b}]{Dravins97b} Dravins D., Lindegren L., Mezey E., \& Young A.~T.\ 1997b, \pasp, 109, 173 

\bibitem[\protect\citeauthoryear{Dravins et al.}{1998}]{Dravins98} Dravins D., Lindegren L., Mezey E., \& Young A.~T.\ 1998, \pasp, 110, 610 

\bibitem[\protect\citeauthoryear{Eichler \& Beskin}{2001}]{Eichler01} Eichler D., \& Beskin G.\ 2001, Astrobiology, 1, 489

\bibitem[\protect\citeauthoryear{Everett \& Howell}{2001}]{Everett01} Everett M.~E., \& Howell S.~B.\ 2001, \pasp, 113, 1428 

\bibitem[\protect\citeauthoryear{Fox}{1992}]{Fox92} Fox G.~K.\ 1992, \mnras, 258, 533 

\bibitem[\protect\citeauthoryear{Frank et al.}{2007}]{Frank07} Frank S., Bentz M.~C., Stanek K.~Z., Mathur S., Dietrich M., Peterson B.~M., \& Atlee D.~W.\ 2007, \apss, 312, 325 

\bibitem[\protect\citeauthoryear{Gillon et al.}{2009}]{Gillon09} Gillon M., et al.\ 2009, \aap, 496, 259 

\bibitem[\protect\citeauthoryear{Grindlay et al.}{1975}]{Grindlay75} Grindlay, J.~E., Helmken, H.~F., Brown, R.~H., Davis, J., \& Allen, L.~R.\ 1975, \apj, 201, 82 

\bibitem[\protect\citeauthoryear{Gruzinov \& Waxman}{1999}]{Gruzinov99} Gruzinov A., \& Waxman E.\ 1999, \apj, 511, 852 

\bibitem[\protect\citeauthoryear{Hanbury Brown, Davis, \& Allen}{Hanbury Brown et al.}{1967}]{HanburyBrown67} Hanbury Brown, R., Davis, J., \& Allen, L.~R.\ 1967, \mnras, 137, 375 

\bibitem[\protect\citeauthoryear{Hanbury Brown, Davis, \& Allen}{Hanbury Brown et al.}{1969}]{HanburyBrown69} Hanbury Brown, R., Davis, J., \& Allen, L.~R.\ 1969, \mnras, 146, 399 

\bibitem[\protect\citeauthoryear{Hartman et al.}{2005}]{Hartman05} Hartman J.~D., Stanek K.~Z., Gaudi B.~S., Holman M.~J., \& McLeod B.~A.\ 2005, \aj, 130, 2241 

\bibitem[\protect\citeauthoryear{Henry}{1995}]{Henry95} Henry G.~W.\ 1995, Robotic Telescopes.~ Current Capabilities, Present Developments, and Future Prospects for Automated Astronomy, 79, 44 

\bibitem[\protect\citeauthoryear{Henry}{1999}]{Henry99} Henry G.~W.\ 1999, \pasp, 111, 845 

\bibitem[\protect\citeauthoryear{Henry et al.}{2000}]{Henry00} Henry G.~W., Marcy G.~W., Butler R.~P., \& Vogt S.~S.\ 2000, \apj, 529, L41 

\bibitem[\protect\citeauthoryear{Hill et al.}{2006}]{Hill06} Hill G.~J., MacQueen P.~J., Palunas P., Kelz A., Roth M.~M., Gebhardt K., \& Grupp F.\ 2006, New Astron. Rev., 50, 378 

\bibitem[\protect\citeauthoryear{Hinton}{2004}]{Hinton04} Hinton J.~A.\ 2004, New Astron. Rev., 48, 331 

\bibitem[\protect\citeauthoryear{Hinton et al.}{2006}]{Hinton06} Hinton J., Hermann G., Kr{\"o}tz P., \& Funk S.\ 2006, Astroparticle Physics, 26, 22 

\bibitem[\protect\citeauthoryear{Holder et al.}{2005}]{Holder05} Holder J., Ashworth P., LeBohec S., Rose H.~J., \& Weekes T.~C.\ 2005, International Cosmic Ray Conference, 5, 387 

\bibitem[\protect\citeauthoryear{Holman \& Murray}{2005}]{Holman05} Holman, M.~J., \& Murray, N.~W.\ 2005, Science, 307, 1288 

\bibitem[\protect\citeauthoryear{Hough et al.}{2006}]{Hough06} Hough J.~H., Lucas P.~W., Bailey J.~A., Tamura M., Hirst E., Harrison D., \& Bartholomew-Biggs M.\ 2006, \pasp, 118, 1302 

\bibitem[\protect\citeauthoryear{Humensky}{2007}]{Humensky07} Humensky T.~B.\ 2007, \apss, 311, 353 

\bibitem[\protect\citeauthoryear{Huovelin \& Piirola}{1990}]{Huovelin90} Huovelin J., \& Piirola V.\ 1990, \aap, 231, 588 

\bibitem[\protect\citeauthoryear{Kawachi et al.}{2001}]{Kawachi01} Kawachi A., et al.\ 2001, Astroparticle Physics, 14, 261 

\bibitem[\protect\citeauthoryear{Kerins et al.}{2010}]{Kerins10} Kerins E., Darnley M.~J., Duke J.~P., Gould A., Han C., Newsam A., Park B.~G., \& Street R.\ 2010, \mnras, 409, 247 

\bibitem[\protect\citeauthoryear{Kildea et al.}{2007}]{Kildea07} Kildea J., et al.\ 2007, Astroparticle Physics, 28, 182 

\bibitem[\protect\citeauthoryear{Krisciunas \& Schaefer}{1991}]{Krisciunas91} Krisciunas K., \& Schaefer B.~E.\ 1991, \pasp, 103, 1033 

\bibitem[\protect\citeauthoryear{Leinert et al.}{1998}]{Leinert98} Leinert C., et al.\ 1998, \aaps, 127, 1 

\bibitem[\protect\citeauthoryear{Le Bohec \& Holder}{2006}]{LeBohec06} Le Bohec S., \& Holder J.\ 2006, \apj, 649, 399

\bibitem[\protect\citeauthoryear{Lewis}{1990}]{Lewis90} Lewis D.~A.\ 1990, Experimental Astronomy, 1, 213 

\bibitem[\protect\citeauthoryear{Loeb \& Gaudi}{2003}]{Loeb03} Loeb A., \& Gaudi B.~S.\ 2003, \apj, 588, L117 

\bibitem[\protect\citeauthoryear{L{\'o}pez-Morales}{2006}]{LopezMorales06} L{\'o}pez-Morales M.\ 2006, \pasp, 118, 716 

\bibitem[\protect\citeauthoryear{Lorenz}{2004}]{Lorenz04} Lorenz E.\ 2004, New Astronomy Reviews, 48, 339 

\bibitem[\protect\citeauthoryear{Lucarelli et al.}{2008}]{Lucarelli08} Lucarelli F., et al.\ 2008, Nuclear Instruments and Methods in Physics Research A, 589, 415 

\bibitem[\protect\citeauthoryear{Miralda-Escud{\'e}}{2002}]{MiraldaEscude02} Miralda-Escud{\'e} 
J.\ 2002, \apj, 564, 1019 

\bibitem[\protect\citeauthoryear{Mirzoyan \& Andersen}{2009}]{Mirzoyan09} Mirzoyan R., \& Andersen, M.~I.\ 2009, Astroparticle Physics, 31, 1 

\bibitem[\protect\citeauthoryear{Mundell et al.}{2007}]{Mundell07} Mundell C.~G., et al.\ 2007, Science, 315, 1822 

\bibitem[\protect\citeauthoryear{Nightingale \& Buscher}{1991}]{Nightingale91} Nightingale N.~S., \& Buscher D.~F.\ 1991, \mnras, 251, 155 

\bibitem[\protect\citeauthoryear{O'Donoghue}{1995}]{ODonoghue95} O'Donoghue D.\ 1995, Baltic Astronomy, 4, 519 

\bibitem[\protect\citeauthoryear{O{\~n}a-Wilhelmi et al.}{2004}]{OnaWilhelmi04} O{\~n}a-Wilhelmi E., Cortina J., de Jager O.~C., \& Fonseca V.\ 2004, Astroparticle Physics, 22, 95 

\bibitem[\protect\citeauthoryear{Ong}{1998}]{Ong98} Ong, R.~A.\ 1998, Physics Reports, 305, 93 

\bibitem[\protect\citeauthoryear{Oosterbroek et al.}{2006}]{Oosterbroek06} Oosterbroek T., de Bruijne J.~H.~J., Martin D., Verhoeve P., Perryman M.~A.~C., Erd C., \& Schulz R.\ 2006, \aap, 456, 283 

\bibitem[\protect\citeauthoryear{Pepper, Gould, \& Depoy}{Pepper et al.}{2003}]{Pepper03} Pepper J., Gould A., \& Depoy D.~L.\ 2003, Acta Astronomica, 53, 213 

\bibitem[\protect\citeauthoryear{Perryman et al.}{1999}]{Perryman99} Perryman M.~A.~C., Favata F., Peacock A., Rando N., \& Taylor B.~G.\ 1999, \aap, 346, L30 

\bibitem[\protect\citeauthoryear{Pont, Zucker, \& Queloz}{Pont et al.}{2006}]{Pont06} Pont F., Zucker S., \& Queloz D.\ 2006, \mnras, 373, 231 

\bibitem[\protect\citeauthoryear{Preu\ss\ et al.}{2002}]{Preu02} Preu\ss\ S., Hermann G., Hofmann W., \& Kohnle A.\ 2002, Nuclear Instruments and Methods in Physics Research A, 481, 229 

\bibitem[\protect\citeauthoryear{Racusin et al.}{2008}]{Racusin08} Racusin J.~L., et al.\ 2008, \nat, 455, 183 

\bibitem[\protect\citeauthoryear{Renker}{2007}]{Renker07} Renker, D.\ 2007, Nuclear Instruments and Methods in Physics Research A, 571, 1 

\bibitem[\protect\citeauthoryear{Reynolds et al.}{2005}]{Reynolds05} Reynolds A.~P., Ramsay G., de Bruijne J.~H.~J., Perryman M.~A.~C., Cropper M., Bridge C.~M., \& Peacock A.\ 2005, \aap, 435, 225 

\bibitem[\protect\citeauthoryear{Richichi}{1994}]{Richichi94} Richichi, A.\ 1994, in Robertson J. G., \& Tango W. J., eds, Very High Angular Resolution Imaging, 158, 71 

\bibitem[\protect\citeauthoryear{Romani et al.}{1999}]{Romani99} Romani R.~W., Miller A.~J., Cabrera B., Figueroa-Feliciano E., \& Nam S.~W.\ 1999, \apj, 521, L153 

\bibitem[\protect\citeauthoryear{Romani et al.}{2001}]{Romani01} Romani R.~W., Miller A.~J., Cabrera B., Nam S.~W., \& Martinis J.~M.\ 2001, \apj, 563, 221 

\bibitem[\protect\citeauthoryear{Roques, Moncuquet, \& Sicardy}{Roques et al.}{1987}]{Roques87} Roques F., Moncuquet M., \& Sicardy B.\ 1987, \aj, 93, 1549 

\bibitem[\protect\citeauthoryear{Roques \& Moncuquet}{2000}]{Roques00} Roques, F., \& Moncuquet, M.\ 2000, Icarus, 147, 530 

\bibitem[\protect\citeauthoryear{Sanchez Almeida \& Martinez Pillet}{1992}]{SanchezAlmeida92} Sanchez Almeida J., \& Martinez Pillet V.\ 1992, \aap, 260, 543 

\bibitem[\protect\citeauthoryear{Schaefer, King, \& Deliyannis}{Schaefer et al.}{2000}]{Schaefer00} Schaefer B.~E., King J.~R., \& Deliyannis C.~P.\ 2000, \apj, 529, 1026 

\bibitem[\protect\citeauthoryear{Schlichting et al.}{2009}]{Schlichting09} Schlichting, H.~E., Ofek, E.~O., Wenz, M., Sari, R., Gal-Yam, A., Livio, M., Nelan, E., \& Zucker, S.\ 2009, \nat, 462, 895 

\bibitem[\protect\citeauthoryear{Schroeder}{2000}]{Schroeder00} Schroeder D.~J.\ 2000, \emph{Astronomical optics}, San Diego : Academic Press, 2000.

\bibitem[\protect\citeauthoryear{Seager, Whitney, \& Sasselov}{Seager et al.}{2000}]{Seager00} Seager S., Whitney B.~A., \& Sasselov D.~D.\ 2000, \apj, 540, 504 

\bibitem[\protect\citeauthoryear{Southworth et al.}{2009}]{Southworth09} Southworth J., et al.\ 2009, \mnras, 396, 1023 

\bibitem[\protect\citeauthoryear{Tinbergen}{1982}]{Tinbergen82} Tinbergen J.\ 1982, \aap, 105, 53 

\bibitem[\protect\citeauthoryear{Verhoeve et al.}{2006}]{Verhoeve06} Verhoeve P., Martin D.~D.~E., Hijmering R.~A., Verveer J., van Dordrecht A., Sirbi G., Oosterbroek T., \& Peacock A.\ 2006, Nuclear Instruments and Methods in Physics Research A, 559, 598 

\bibitem[\protect\citeauthoryear{Verhoeve}{2008}]{Verhoeve08} Verhoeve P.\ 2008, Journal of Low Temperature Physics, 151, 675 

\bibitem[\protect\citeauthoryear{Weekes et al.}{2002}]{Weekes02} Weekes T.~C., et al.\ 2002, Astroparticle Physics, 17, 221 

\bibitem[\protect\citeauthoryear{Wiktorowicz}{2009}]{Wiktorowicz09} Wiktorowicz S.~J.\ 2009, \apj, 696, 1116 

\bibitem[\protect\citeauthoryear{Wiktorowicz \& Matthews}{2008}]{Wiktorowicz08} Wiktorowicz S.~J., \& Matthews K.\ 2008, \pasp, 120, 1282 

\bibitem[\protect\citeauthoryear{Winn, Holman, \& Roussanova}{Winn et al.}{2007}]{Winn07} Winn, J.~N., Holman, M.~J., \& Roussanova, A.\ 2007, \apj, 657, 1098 

\bibitem[\protect\citeauthoryear{Winn et al.}{2009}]{Winn09} Winn, J.~N., Holman, M.~J., Carter, J.~A., Torres, G., Osip, D.~J., \& Beatty, T.\ 2009, \aj, 137, 3826 

\bibitem[\protect\citeauthoryear{Young}{1967}]{Young67} Young A.~T.\ 1967, \aj, 72, 747 

\bibitem[\protect\citeauthoryear{Zucker, Mazeh, \& Alexander}{Zucker et al.}{2007}]{Zucker07} Zucker S., Mazeh T., \& Alexander T.\ 2007, \apj, 670, 1326 



\end{thebibliography}
\end{document}